\documentclass[prd,superscriptaddress,11pt,nofootinbib]{revtex4-1}
\usepackage{amsmath}    
\usepackage[english]{babel}
\usepackage[utf8]{inputenc}
\usepackage{graphicx}   
\usepackage{slashed}
\usepackage{epstopdf}
\usepackage{verbatim}   
\usepackage{color}      
\usepackage{subfigure}  
\usepackage{multirow}
\usepackage{hyperref}   
\usepackage{float}
\usepackage{axodraw}
\restylefloat{table}
\usepackage{bm}
\newcommand{\be}{\begin{eqnarray}}
\newcommand{\ee}{\end{eqnarray}}
\newcommand{\Lm}{\mathcal{L}}
\newcommand{\h}{{\mathcal H}}
\newcommand{\eps}{\epsilon}

\def\addresses#1#2{\hbox to \hsize{\@tablebox{#1}\hfil\@tablebox{#2}}}
\def\@tablebox#1{\vtop{\hsize=5in \begin{flushleft} #1 \end{flushleft}}}

\def\beq{\begin{equation}}
\def\eeq{\end{equation}}
\def\bit{\begin{itemize}}
\def\eit{\end{itemize}}
\def\beqa{\begin{eqnarray}}
\def\eeqa{\end{eqnarray}}
\def\bray{\begin{array}}
\def\eray{\end{array}}

\def\n{\nonumber}

\def\zb{\bar z}

\def\fr{\frac}
\def\_{\textunderscore}

\def\M{\mathcal{M}}

\definecolor{orange}{rgb}{1,0.5,0}

\begin{document}
\author{Junmou Chen}
\email{jmchen@kias.re.kr}
\affiliation{School of Physics, Korean Institute for Advanced Study, Seoul, 02455, Korea}
\title{{\bf On the  Feynman Rules  of Massive Gauge Theory in Physical Gauges }}

\begin{abstract}
For a massive gauge theory with Higgs mechanism in a physical gauge, the longitudinal polarization of gauge bosons can be naturally identified as  mixture of the goldstone component and a remnant gauge component that vanishes at the limit of zero mass, making the goldstone equivalence manifest.  In light of this observation, we re-examine the Feynman rules of massive gauge theory by treating gauge fields and their corresponding goldstone fields as single objects, writing them uniformly as 5-component ``vector" fields.  The gauge group is taken to be $SU(2)_L$ to preserve custodial symmetry.  We find the derivation of gauge-goldstone propagators becomes rather trivial by noticing there is a remarkable parallel between massless gauge theory and massive gauge theory in this notation.  We also derive the Feynman rules of all vertices, finding the vertex for self-interactions of vector (gauge-goldstone) bosons are especially simplified. 
We then demonstrate that the new form of the longitudinal polarization vector and the standard form give the same results for all the 3-point on-shell amplitudes.  This on-shell matching confirms similar results obtained with on-shell approach for massive scattering amplitudes by Arkani-Hamed et.al. in ref.(\cite{Arkani-Hamed:2017jhn}).  Finally we calculate some $1\rightarrow 2$ collinear splitting amplitudes by making use of the new Feynman rules and the on-shell match condition.  
\end{abstract}

\keywords{}

\maketitle

\section{Introduction}









In a massive gauge theory with Higgs mechanism, scattering amplitudes involving longitudinal polarizations have the famous  problem of power counting\cite{Lee:1977eg,Lee:1977yc,Cornwall:1974km}: while single Feynman diagrams increase with energy, the S-matrix is well-behaved when taking into account the contribution of the Higgs boson.  This failure of power counting causes many complications and confusion both practically and conceptually.  The origin of the problem is the longitudinal polarization vector behave as $\epsilon_L^{\mu} \sim \frac{k^{\mu}}{m_W}+O(\frac{m_W}{E})$ in high energy limit. Another way to phrase it is that the longitudinal polarization vector and Feynman diagrams don't have a smooth limit as $m_W\rightarrow 0$, thus it's not clear how the theory approaches massless limit continuously. In  future high energy colliders \cite{Arkani-Hamed:2015vfh}\cite{Mangano:2017tke}, we will approach energy scales in which the EW symmetry will be effectively restored. This problem becomes even more severe.

Practically this problem is often solved by replacing the longitudinal vector bosons with the corresponding goldstone bosons, according to the so-called  goldstone equivalence theorem(GET)\cite{Chanowitz:1985hj,Gounaris:1986cr,Bagger:1989fc,Veltman:1989ud,Yao:1988aj}, which states that scattering amplitudes involving longitudinal vector bosons  can be approximated by the corresponding goldstone modes in high energy limit:

\be\label{eq:GET}
{\mathcal M}(W_L,W_L, W_L, ...., W_L)= (-i)^n{\mathcal M}(\phi, \phi, \phi, ...., \phi)+ O(\frac{m_W}{\sqrt{s}}), 
\ee

\

\noindent with $\sqrt{s}$ being the hard scale of the process. However, this solution is still not completely satisfactory, as 
GET is only an approximation with other terms suppressed in high energy limit. Although the approximation seems to work for naive power counting, it's not utterly clear if the contributions of  the terms neglected by GET are real subdominant.  In fact,  it was discovered in \cite{Chen:2016wkt} that there is a new class of splitting functions contributing to DGLAP evolution of EW PDFs and substructure of EW jets.  Those new splitting functions originate precisely from the terms that are neglected by GET.\footnote{The mistake of the naive power counting is that it neglects that a physical process is intrinsically multi-scaled. The terms neglected by GET has soft singularities (with infrared cut-offs provided by the masses) that give rise to contributions up to single logarithms, when the collinear scale $\lambda$ lies in  $ m_W\ll \lambda \ll \sqrt{s}$.    }   It then becomes mandatory to account for all the terms that from the longitudinal polarization vector in calculation.  Obviously we need a better solution for the power counting problem.

A physical gauge in a massive gauge theory can be defined by the gauge-fixing condition $n\cdot W=0$, with $n$ being any direction other than $k$\footnote{Traditionally a physical gauge is defined for $n^{\mu}$ along a fixed direction, e.g. $n^{\mu}=(1,0,0,1)^T$.  Here we adopt a more general definition, which includes $n$  being momentum dependent, e.g. Coulomb gauge\cite{Beenakker:2001kf}. Other examples of momentum dependent ``physical" gauge can be found in \cite{Chen:2016wkt} and \cite{Feige:2014wja}. }, is able to serve this purpose~\cite{Chen:2016wkt,Kunszt:1987tk,Borel:2012by,Beenakker:2001kf}.  Heuristically we can argue this way: GET is the consequence of gauge symmetry. It can be derived from Ward identities of the theory. Nevertheless,  there is also an alternative and a more direct way to prove GET, i.e. we can simply choose another gauge.  Since in a physical gauge we only impose gauge-fixing on gauge fields, the gauge-goldstone mixing term in the Lagrangian remains, thus  we are forced to identify gauge fields $W^{\mu}$ and goldstone fields $\phi$ as single fields, which we can denote as $W^M=(W^{\mu}, \phi)$. In the resulting gauge-goldstone propagator,  goldstone modes and gauge modes obtain the same pole masses, the longitudinal polarization vectors are naturally identified  as  mixture of  gauge components and goldstone components. We can write the longitudinal polarization vector as $\epsilon_L^M=(\epsilon_n^{\mu}, -i)$, with $\epsilon_n^{\mu}\sim -\frac{m_W}{E}n^{\mu}$ in high energy limit. Its specific form depends on the gauge direction $n$.  In this way, we obtain a precise formula of scattering amplitudes involving longitudinal vector bosons, which is a generalization of GET in Eq.(\ref{eq:GET}),  
 
 \be\label{eq:GET_2}
{\mathcal M}(W_L,W_L,W_L, ...., W_L) &=& (-i)^n{\mathcal M}(\phi, \phi, \phi, ...., \phi)+ (-i)^{n-1}{\mathcal M}(W_n, \phi, \phi, ...., \phi)  \n \\
&&+ .... + {\mathcal M}(W_n,W_n,W_n, ...., W_n)
\ee

\noindent The polarization vectors of $W_n$ are given by $\epsilon_n$, which is usually neglected by GET.  
Thus physical gauges are vastly different from $R_\xi$ gauge,  in which the masses of the goldstone bosons are gauge-dependent.  Of course, physical results cannot be gauge-dependent.  The author in  \cite{Wulzer:2013mza}  obtained similar results as Eq.(\ref{eq:GET_2}) based on  Feynman gauge by making  use of BRST symmetry to redefine the physical state.  An earlier attempt along this line can be found in \cite{Espriu:1994ep}.  The longitudinal polarization and related scattering amplitudes in physical gauges agree precisely with those in $R_\xi$ gauge in \cite{Wulzer:2013mza}\cite{Espriu:1994ep} if the gauge direction is chosen as $n^{\mu}=(1,-\hat{k})$, with $\hat{k}=\frac{|\vec{k}|}{\vec{k}}$.  Thus the two approaches are equivalent with each other.  Nevertheless,  comparing to $R_\xi$ gauge, physical gauges provide a much more clear physical picture as there is no gauge-dependent goldstone mass,  no  ambiguity in identifying physical states through LSZ reduction formula. 

Although the power counting problem is overcome in a physical gauge, there is also a drawback:  the Feynman rules become complicated, as we need to sum over all the terms from both gauge components and goldstone components.  The problem becomes especially severe if the number of  longitudinal states are multiplied.  Besides, the derivation of the gauge-goldstone propagators seems also to be complicated due to the gauge-goldstone mixing terms in the Lagrangian.  The goal of this paper  is to investigate and reorganize the Feynman rules of massive gauges in physical gauges by combining gauge fields and goldstone fields together.   The  strategy is, as mentioned above, to treat gauge fields and goldstone fields uniformly as 5-component fields $W^M=(W^{\mu}, \phi)$, and exploit possible underlying structures to simplify.   The model we choose is the $\theta_W\rightarrow 0$ limit of the Standard Model of EW interactions, so the gauge group is $SU(2)_L$ instead of $SU(2)_L\times U(1)$.  The motivation is that  if custodial symmetry is preserved, both the gauge fields and goldstone fields transform as a triplet under $SU(2)$ global symmetry.  It then becomes straightforwardly to combine gauge components and goldstone components.   Additionally, it's noteworthy that the new Feynman rules can also apply to  Feynman gauge, in which goldstone bosons obtain the same masses as their corresponding gauge bosons. So we can make use of the Feynman rules describe here, if the longitudinal polarization vector is taken to be $\epsilon_L^M=(\epsilon_n^{\mu}, -i)$.

 Apart from deriving and documenting the Feynman rules, we also investigate all the 3-point on-shell amplitudes. In recent years, the on-shell approach of scattering amplitudes using spinor-helicity\cite{Dixon:2013uaa}\cite{Elvang:2013cua} has made remarkable progress. However, the success is still largely confined in massless particles. There have been many papers\cite{Arkani-Hamed:2017jhn,Badger:2005jv,Badger:2005zh,Craig:2011ws} devoted to the massive case, but the topic still remains largely unexplored. 3-point on-shell amplitudes are the building blocks of on-shell approach to scattering amplitudes, thus one might hope  that clearer understanding of them can shed some light in the direction.  Our basic point is, now that we have two forms of longitudinal polarization vectors -- one from gauge fields only, another from mixture of gauge fields and goldstone fields -- the two forms should give the same amplitudes due to gauge invariance.  This match between two ways of evaluating amplitudes should also be reflected on the 3-point on-shell amplitudes, which can be appropriately called ``on-shell match".  This on-shell match gives a way to explain how the information of goldstone bosons are ``encoded" in gauge fields for the case of 3-point amplitudes. 
 
 Another motivation for 3-point on-shell amplitudes  is the calculation of collinear splitting functions, which can be reduced to the calculation of $1\rightarrow 2$ collinear splitting amplitudes.  Since collinear singularity emerges as the internal lines of the Feynman diagrams approach the mass poles,  collinear splitting amplitudes are simply on-shell amplitudes.  In light of this observation, we can simplify the calculation of splitting amplitudes -- especially for massive particles -- by making use of the new Feynman rules and the on-shell match condition.


The remaining of this paper is organised as following: 

 In Section \ref{sec:derivation} we write down the Lagrangian of the model, derive all the Feynman rules in physical gauges.   We first derive the propagators and polarizations, we then derive all the vertices systematically.  All the Feynman rules at tree level are listed  in  the appendix \ref{sec:feynrules}.
 
In Section \ref{sec:3-point} we investigate all 3-point on-shell amplitudes. We first prove all those 3-point amplitudes satisfy on-shell gauge symmetry, then calculate collinear splitting amplitudes involving longitudinal vector bosons by making use of  the Feynman rules obtained in this paper and on-shell match condition from on-shell gauge symmetry.  Finally we have conclusions.

\section{The Model and Feynman Rules}
\label{sec:derivation}

\subsection{Lagrangian}
\label{sec:lag}

Our goal is to derive the Feynman rules of the Standard Model of  Electroweak interactions  by taking 
 the $\theta_W\rightarrow 0$ limit. The  gauge  group is thus  $SU(2)_L$ only.  With the custodial symmetry, the Higgs potential has symmetry $SU(2)_L\times SU(2)_R$.  We can parametrize the Higgs field  by writing it as  

\[\h=\frac{1}{\sqrt{2}}(i\sigma_2\Phi^*, \Phi)\]

\noindent with  $\Phi$ being 
 
\[ 
\Phi = 
\left(\begin{array}{c}  
\frac{1}{\sqrt{2}} -i(\phi_1-i\phi_2)\\
 h+i\phi_0
\end{array}
\right)
 \]



A more illuminating way to  write  Higgs doublet $\h$ is 

\be\label{eq:higgs}
\h =\frac{1}{\sqrt{2}} (i\sigma_2\Phi^*, \Phi) = \frac{1}{2}(h -i \sigma^a\phi_a)
\ee

\indent The would-be goldstone fields are isolated from the would-be Higgs field in this parametrization, which will be  more convenient to treat the would-be goldstone bosons  as the 5th component of the vector fields/states. The full Lagrangians are written as

\be\label{eq:lagrangian}
{\mathcal L}_{\text{Gauge}} &=& -\frac{1}{4} (W_{\mu\nu}^aW^{\mu\nu}_a) +\frac{1}{2\xi} (n\cdot \partial \ n\cdot W^a)(n\cdot \partial \ n\cdot W_a)^* \n  \\
{\mathcal L}_{\text{Higgs}}&=&\text{tr}[(D_{\mu}\h)^{\dagger}D^{\mu}\h]-\frac{\lambda_h}{4}({\text Tr}[\h^{\dagger}\h]-\frac{v^2}{2})^2 \\
{\mathcal L}_{\text{Fermion}} &=& i\sum_{i=1,2,3}\overline{Q}'^i_L\slashed{D}Q'^i_L + i\sum_{i=1,2,3}\overline{L}'^i_L\slashed{D}L'^i_L \n \\
{\mathcal L}_{\text{Yukawa}} &=& -\sum_{ij=1,2,3}\sqrt{2}\overline{Q}'^i_L\h {\bf Y}_{Q}^{'ij}Q'^j_R -\sum_{i,j=1,2,3}\sqrt{2}\overline{L}'^i_L\h {\bf Y}_{L}^{'ij}L'^j_R+ \text{h.c.} \n
\ee

Here $W_{\mu\nu}^a=\partial_{\mu}W_{\nu}^a-\partial_{\nu}W_{\mu}^a-g\epsilon^{abc}W_{\mu}^bW_{\nu}^c$, $D_{\mu}=\partial_{\mu}+ig\frac{\sigma^a}{2}W^{a}_{\mu}$,  $n^{\mu}$ can be either a fixed direction\cite{Kunszt:1987tk}\cite{Borel:2012by} (e.g. $n^{\mu}=(1,0,0,1)$), or an operator depending on the coordinates\cite{Chen:2016wkt} \cite{Beenakker:2001kf}. To make sure gauge-fixing parameter $\xi$ is dimensionless,  $n^{\mu}$ is  rescaled by $n\cdot \partial$. Ghosts in physical gauges generally decouple from the theory, but  don't decouple if $n$ is momentum-dependent \cite{Feige:2014wja}.  Nevertheless, we restrict our focus on tree level in this paper.  

For the Fermion sector and the Yukawa sector,     $Q'^i_{L/R}=(u'^i_{L/R}, d'^i_{L/R})$ and  $L'^i_{L/R}=(\nu'^i_{L/R}, e'^i_{L/R})$ denote quarks and leptons in flavor basis respectively. Indices $i,j=1,2,3$ denotes generations.  ${\bf Y}^{'ij}_{Q}= \text{diag}(y_u^{\prime ij}, y_d^{\prime ij})$ is Yukawa matrix for the quark sector in isospin space,${\bf Y}^{'ij}_{L}= \text{diag}(y_\nu^{\prime ij}, y_l^{\prime ij})$ is Yukawa matrix for the lepton sector in isospin space. 
\

  The Lagrangian terms in Eq.(\ref{eq:lagrangian}) are invariant under  

\be
\h \rightarrow e^{\frac{i\alpha^a\sigma_a}{2}}\h   \n \\
    Q_L \rightarrow e^{\frac{i\alpha^a\sigma_a}{2}}Q_L  \\
     L_L \rightarrow e^{\frac{i\alpha^a\sigma_a}{2}}L_L \n
\ee


Next we expand the Lagrangian terms in terms of $W_{\mu}$, $h$, $\phi$.  After symmetry breaking, the Higgs field has shift: $h\rightarrow h +v$.  Particles obtain masses, the relations between masses and $v$ are 

\be
m_W=\frac{gv}{2} \ \  \  m_f=\frac{y_f v}{\sqrt{2}}\  \  \  m_h^2=\frac{\lambda_h v^2}{2}
\ee

We start with gauge sector and Higgs sector.  For the gauge sector, the Lagrangian terms can be written as 

\be\label{eq:vertices_1}
{\mathcal L}_{W_{\mu}^2} &=&-\frac{1}{2}\partial^{\mu}W^a\partial_{\mu}W_a+\frac{1}{2\xi} (n\cdot \partial \ n\cdot W^a)(n\cdot \partial \ n\cdot W_a)^* \n \\
{\mathcal L}_{W_{\mu}^3}+{\mathcal L}_{W^4_{\mu}}&=& g\epsilon^{abc}\partial^{\mu}W^{\nu a}W_{\mu}^bW_{\nu}^c-\frac{g^2}{4}\epsilon^{abc}\epsilon^{afg}W_{\mu}^bW_{\nu}^cW^{\mu f}W^{\nu g}.
\ee

For the Higgs sector,   the covariant derivative on the Higgs doublet  $\h$ is written as 

\be
D_{\mu}\h = (\partial_{\mu}+igW_{a\mu}\frac{\sigma^a}{2}) ((h+v)\frac{{\textbf 1}}{2}-i\frac{\sigma^b}{2}\phi_b) \n
\ee

Making use of  $\sigma^a\sigma^b=\delta^{ab}{\textbf 1}+i\epsilon^{abc}\sigma^c$ and separating $h$ and $\phi$,   $D_{\mu}\h$ can be written further as 

\be 
       D_{\mu}\h          = (\partial_{\mu}+igW_{a\mu}\frac{\sigma^a}{2})\cdot (h+v)\frac{{\textbf 1}}{2} 
             -i\frac{\sigma^a}{2}(\partial_{\mu}\delta^{ac}-\frac{g}{2}\epsilon^{abc}W^b_{\mu})\phi^c+\frac{1}{4}gW_{\mu}^a\phi_a{\textbf 1}.
\ee
             

Then we plug in   $D_{\mu}\h$ into $\text{tr}[(D^{\mu}\h)^{\dag}D_{\mu}\h]$. Combined with  the Higgs potential $V(\text{Tr}(\h^{\dag}\h))$, the Lagrangian terms for Higgs sector become

\be\label{eq:vertices_2}
{\mathcal L}_{h^2} &=& \frac{1}{2}\partial^{\mu}h\partial_{\mu}h-\frac{1}{2}m_h^2h^2
\n \\ 
{\mathcal L}_{\phi^2+\phi W_{\mu}^2+\phi^2W_{\mu}^2} &=& \frac{1}{2}\partial^{\mu}\phi^a\partial_{\mu}\phi_a -\frac{g}{2}\epsilon^{abc}\partial^{\mu}\phi^aW_{\mu}^b\phi^c
           +  \frac{g^2}{8}\epsilon^{abc}\epsilon^{afg}W_{\mu}^b\phi^cW^{f\mu}\phi^g + \frac{g^2}{8}W^{a\mu}\phi_aW^{b}_{\mu}\phi_b.
\ee

and 

\be\label{eq:vertices_3}
{\mathcal L}_{\phi W_{\mu}+hW_{\mu}^2+h\phi W_{\mu}}&=& -m_W\partial_{\mu}\phi^aW^{\mu}_a + (\frac{g^2}{8}h^2+\frac{gm_W}{2}h + \frac{m_W^2}{2})W^{\mu}_aW_{\mu}^a+\frac{g}{2}(\partial_{\mu}hW^{a\mu}\phi_a-\partial_{\mu}\phi^aW^{\mu}_ah) \n \\
{\mathcal L}_{h^3+h^4}&=&-\frac{1}{16}\lambda_hh^4-\frac{1}{4}\lambda_hvh^3\n\\
{\mathcal L}_{h^2\phi^2}&=&-\frac{1}{8}\lambda_hh^2\phi^a\phi_a-\frac{1}{4}\lambda_hvh\phi^a\phi_a \\
{\mathcal L}_{\phi^4}&=&-\frac{1}{16}\lambda_h\phi^a\phi_a\phi^b\phi_b\n
\ee






For the fermion sector,  after symmetry breaking, $Q^{\prime i}$ and $L^{\prime i}$ are related to the mass basis by 

\be
Q^{\prime i} = U_Q^{ij} Q^j = \left( \begin{array}{cc}
U_u^{ij} & 0 \\
0 &  U_d^{ij}
\end{array}  \right)  \left(\begin{array}{c} 
u^j \\
d^j
 \end{array}\right)
 \  \   \   L^{\prime i} = U_L^{ij} L^j = \left( \begin{array}{cc}
U_\nu^{ij} & 0 \\
0 &  U_l^{ij}
\end{array}  \right)  \left(\begin{array}{c} 
\nu^j \\
l^j
 \end{array}\right)
\ee

\noindent The Yukawa matrices and mass matrices are diagonalized  by  the mixing matrices $U_{Q/L}$, 

\be
{\bf Y}_{Q/L}^{il}= U^{\dag ij}_{Q/L}{\bf Y}^{\prime jk}U^{kl}_{Q/L} = {\bf Y}^{i}_{Q/L}\delta_{il}
\ee

\noindent as well as 

\be
{\bf m}_{Q/L}^{il}= U^{\dag ij}_{Q/L}{\bf Y}^{\prime jk}U^{kl}_{Q/L}\frac{v}{\sqrt{2}} = {\bf m}^{i}_{Q/L}\delta_{il}
\ee

\noindent  with ${\bf Y}^i_{Q}=\text{diag}(y_{u_i}, y_{d_i})$,  ${\bf Y}^i_L=\text{diag}(y_{\nu_i}, y_{l_i})$,  ${\bf m}_{Q}^i=\text{diag}(m_{u_i}, m_{d_i})$,  ${\bf m}^i_L=\text{diag}(m_{\nu_i}, m_{l_i})$. 
The Lagrangian terms for the fermion sector then become  

\be\label{eq:vertices_30}
{\mathcal L}_{f^2}&=& i\overline{Q}_L\slashed{\partial}Q_L + i\overline{L}_L\slashed{\partial}L_L-\overline{Q}_L{\bf m}_QQ_R -\overline{L}_L{\bf m}_LL_R  \n \\
{\mathcal L}_{ffh}&=&-\frac{1}{\sqrt{2}} \overline{Q}_L{\bf Y}_QQ_R h -\left(\frac{1}{\sqrt{2}} \overline{L}_L{\bf Y}_LL_R h + \text{h.c.} \right) \n \\
{\mathcal L}_{ffW_{\mu}}+ {\mathcal L}_{ff\phi}&=& -\frac{g}{2}\overline{Q}_L\gamma^{\mu}(U^\dag_Q\sigma^aU_Q)Q_L W_{\mu a}+\left(\frac{i}{\sqrt{2}} \overline{Q}_L(U^\dag_Q\sigma^a{\bf Y}_QU_Q)Q_R\phi^a  + \text{h.c.}\right)  \\
&& -\frac{g}{2}\overline{L}_L\gamma^{\mu}(U^\dag_L\sigma^aU_L)L_L W_{\mu a}+\left(\frac{i}{\sqrt{2}} \overline{L}_L(U^\dag_L\sigma^a{\bf Y}_LU_L)L_R\phi^a + \text{h.c.} \right) \n
\ee

\noindent The generation indices have been suppressed. 



     
 


\subsection{Propagator}
\label{eq:prop}


In this section we are deriving the propagator of vector bosons,  which has intrinsic mixing between gauge modes and goldstone modes.  Combining the kinetic terms in Eq.(\ref{eq:vertices_1}), Eq.(\ref{eq:vertices_2}) and Eq.(\ref{eq:vertices_3}), the quadratic Lagrangian terms for gauge fields and goldstone fields are

\be\label{eq:kinetic}
\Lm_{W_a^2}&=&-\frac{1}{2}\partial^{\mu}W_a^{\nu}\partial_{\mu}W_{a\nu}+\frac{1}{2}\partial^{\mu}W_{a\mu}\partial^{\nu}W_{a\nu}+\frac{1}{2}m_W^2W_{a\mu}W^{a\mu} \nonumber \\
& +&\frac{1}{2\xi}(n\cdot \partial  \ n\cdot W_{a})(n\cdot \partial \ n \cdot W_{a})^*  \nonumber \\
\Lm_{\phi_aW^a}&=&-m_WW^{a\mu}\partial_{\mu}\phi_a\\
\Lm_{\phi_a^2}&=&\frac{1}{2}(\partial^{\mu}\phi_a)^2  \nonumber
\ee

We note that Eq.(\ref{eq:kinetic})  is not only true for the $SU(2)_L$ theory, but applies to any model with Higgs mechanism. 
We write gauge-goldstone fields as 5-component vector fields $W^M_a=(W_a^{\mu}, \phi_a)$,  then the kinetic Lagrangian terms can be written as following up to terms with total derivative, 



\be
{\mathcal L}_{W_M^2}=-\frac{1}{2}\partial_MW_N^a\partial^MW^N_a+\frac{1}{2}(\partial_MW_a^{M})^2+\frac{1}{2\xi}(n\cdot \partial \ n_MW_a^M)(n\cdot \partial  \ n_MW_a^M)^*
\ee

with $n^M=(n^{\mu}, 0)$,  $W_M^a=(W_{\mu}^a, \phi^a)$,  $\partial^M=(\partial^{\mu}, -m_W)$, $g_{MN}=g^{MN}=\text{diag}(1, -1, -1, -1, -1)$.  It looks the same as the kinetic Lagrangian terms of massless gauge theory, except $\mu$ becomes $M$. This similarity is not just a nice way of writing all the Lagrangian terms. Indeed,  noticing the Fourier transformation of $\partial^M=(\partial^{\mu}, -m_W)$ gives $k^{M}=(k^{\mu}, -im_W)$ for inwards momentum, and $k^{*M}=(k^{\mu}, im_W)$ for outwards momentum, we could write the dot product of $k^M$ as  

\be
k\cdot k^*=g_{MN}k^Mk^{*N}=k^2-m_W^2
\ee

This equals $0$ when on-shell, just as $k\cdot k =k^2=0$ when on-shell for massless case.  Thus all the algebra with the tensor $g^{\mu\nu}$ and $k^{\mu}$, could be applied  straightforwardly to $g^{MN}$ and $k^M/k^{*N}$, with $n^{M}=(n^{\mu}, 0)$.  For massless gauge fields,  the kinetic  Lagrangian after gauge fixing  is 

\be
{\mathcal L}_{\text{kinetic}}= -\frac{1}{2}\partial_{\mu} W_{\nu}^a\partial^{\mu}W^{\nu}_a+\frac{1}{2}(\partial\cdot W_a)^2+\frac{1}{2\xi}(n\cdot \partial \ n\cdot W_a)(n\cdot \partial  \ n\cdot W_a)^* + \text{total derivative}
\ee

\noindent  the propagator of the gauge bosons can be easily evaluated to be

\be\label{eq:massless_prop}
<W_a^{\mu}W_b^{\nu}> = \frac{-i\delta_{ab}(g^{\mu\nu}-\frac{n^{\mu}k^{\nu}+k^{\mu}n^{\nu}}{n\cdot k}+\frac{n^2k^{\mu}k^{\nu}}{(n\cdot k)^2}+\xi \frac{k^2}{(n\cdot k)^4}k^{\mu}k^{\nu})}{k^2+i\epsilon} 
\ee

Following the arguments above,  a direct analogue to the massless propagator in Eq.(\ref{eq:massless_prop}) gives us the gauge-goldstone propagator in massive gauge theory, 

\be
<W^{M}_aW^{N}_b>=\frac{-i\delta_{ab}(g^{MN}-\frac{n^{M}k^{* N}+k^{M}n^{*N}}{n\cdot k }+\frac{n^2k^Mk^{*N}}{(n\cdot k)^2}+\xi \frac{k\cdot k^*}{(n\cdot k)^4}k^{M}k^{*N})}{k\cdot k^*+i\eps}
\ee



By  writing gauge  components $M=\mu$ and $M=4$ component separately, the propagator of vector boson becomes

\[<(W_a^{\mu},\phi_a), (W_b^{\nu},\phi_b)>=\frac{i\delta_{ab}}{k^2-m_W^2+i\epsilon}\left({
\begin{array}{cc}
-(g^{\mu\nu}-\fr{n^{\mu}k^{\nu}+k^{\mu}n^{\nu}}{n\cdot k}+n^2\frac{k^{\mu}k^{*\nu}}{(n\cdot k)^2})& \ \ \ \ i\fr{m_W}{n\cdot k}(n^{\mu}-n^2\frac{k^{\mu}}{n\cdot k})\\
 -i\fr{m_W}{n\cdot k}(n^{\nu}-n^2\frac{k^{\nu}}{n\cdot k})& 1-\frac{n^2m_W^2}{(n\cdot k)^2}
\end{array}}\right)
\]

When $k^2=m_W^2$ or $k\cdot k^*=0$, the numerator of the propagator can be written as sum of the polarizations, 

\[<W^{M}_aW^{*N}_b>= \frac{i\delta_{ab}\sum_{s=\pm, L}\epsilon_s^M\epsilon_s^{N^*}}{k\cdot k^*+i\eps}\]

In the 5-component notation, the transverse and longitudinal polarizations are

\be\label{eq:pol_1}
 g^{44}=-1:   \  \  \    \    \   \   \ 
\epsilon^{M}_{\pm}= \left(\begin{array}{c}\epsilon_{\pm}^{\mu}\\
       0 \end{array}\right)  \   \   \  
\epsilon^M_L=\frac{1}{\sqrt{1-\frac{n^2m_W^2}{(n\cdot k)^2}}}
\left(\begin{array}{c}
-\frac{m_W}{n\cdot k }(n^{\mu}-\frac{n^2k^{\mu}}{n\cdot k}) \\
 i(1-\frac{n^2m_W^2}{(n\cdot k)^2})
\end{array}\right) 
\ee

They satisfy the transverse condition and normalization condition

\be
\epsilon_s \cdot \eps_{s'}^*&=&-1\cdot \delta_{ss'} \n \\  
k^*\cdot \epsilon_s&=&k\cdot \epsilon_s^*=0 \\
n\cdot \epsilon_{s=\pm}^{(*)} &=&0 \n
\ee









The amplitudes involving longitudinal vector bosons are evaluated by summing over the contributions from both gauge components and goldstone components: 

\be\label{eq:amplitudes}
i{\mathcal M}(L)= ig^{MN}{\mathcal M}_{M}\epsilon_{N}= i{\mathcal M}^{\mu}\epsilon_{n\mu}-i{\mathcal M}^4\epsilon_4
\ee


Notice  the ``metric"  $g^{MN}=\text{dig}(1,-1,-1,-1,-1)$  induces a minus sign between amplitudes involving gauge component and goldstone component.  In practical calculations it's more convenient to  absorb this  minus sign into the polarization vectors, which  become

\be\label{eq:pol_2}
 g^{44}=1:   \  \  \    \    \   \   \ 
\epsilon^M_L=\frac{1}{\sqrt{1-\frac{n^2m_W^2}{(n\cdot k)^2}}}
\left(\begin{array}{c}
-\frac{m_W}{n\cdot k }(n^{\mu}-\frac{n^2k^{\mu}}{n\cdot k}) \\
-i(1-\frac{n^2m_W^2}{(n\cdot k)^2})
\end{array}\right) 
\ee

so that the amplitude is evaluated by simply summing over the diagrams involving gauge components and goldstone components, i.e. 

\be
i{\mathcal M}(L)= i{\mathcal M}^{\mu}\epsilon_{n\mu}+i{\mathcal M}^4\epsilon_4
\ee


Our results agree with Ref.\cite{Chen:2016wkt} by choosing $n^{\mu}=(1,-\frac{\vec{k}}{|\vec{k}|})$ up to a global phase which is unphysical. 

\subsection{ Vertices}
\label{sec:vertices}






In this section we apply the 5-component treatment further to vertices.   Our goal is to recombine the Lagrangian terms for interactions, so that the Feynman rules for vector bosons is given by Lagrangian of  $W^a_M=(W^a_{\mu}, \phi^a)$.   We start from the gauge sector and Higgs sector, which give rise to vertices of   W-W-W and W-W-W-W. The corresponding Lagrangian terms are written as 

\be\label{eq:vertices_10}
{\mathcal L}_{W_M^3}&=&g\epsilon^{abc}\partial_{\mu}W_N^aW_{\rho }^bW_{K}^cg^{\mu\rho}g^{\prime NK}  \\
{\mathcal L}_{W_M^4}&=&-\frac{g^2}{4}\eps^{abc}\eps^{aef}W_{\mu}^bW_{\nu }^eW_{P}^cW_{K}^fg^{\mu\nu}g^{\prime PK}+\frac{g^2}{8}W_{\mu}^aW_{\nu  }^b\phi^a\phi^b g^{\mu\nu}   -\frac{1}{16}\lambda_h\phi^a\phi_a\phi^b\phi_b \n
\ee

with  $g^{\prime NK}=\text{diag}(1, -1, -1, -1, -1/2)$. The $g^{\prime NK}$ here  is not to be confused  with  $g^{NK}$ appearing say in Eq.(\ref{eq:amplitudes}): $g^{\prime NK}$ in Eq.(\ref{eq:vertices_10}) is for bookkeeping the relative coefficient between the different Lagrangian terms, whereas $g^{NK}$ in Eq.(\ref{eq:amplitudes}) is to keep track of the relative phase between (sub)diagrams of gauge components and goldstone components, which can be absorbed into the definition of polarization vectors. 

The Lagrangian terms  for  h-W-W and h-h-W-W  are

\be\label{eq:vertices_20}
{\mathcal L}_{hW_M^2}&=& \frac{g}{2}(\partial_{\mu}hW^a_{\nu}\phi_{a}g^{\mu\nu}-\partial_{\mu}\phi^{a}W_{\nu}^ah g^{\mu\nu}) +\frac{1}{2}gm_WhW_{\mu}^aW_{\nu}^ag^{\mu\nu}-\frac{1}{4}\lambda_hvh\phi^a\phi_a  \\
{\mathcal L}_{h^2W_M^2}&=&\frac{g^2}{8}h^2W_{\mu}^aW_{\nu}^ag^{\mu\nu}-\frac{1}{8}\lambda_hh^2\phi^a\phi_a \n
\ee



\noindent To obtain  Feynman rules we need the last step of writing $W^{a}$  in the basis $(W^{\pm}_M, W^3_M)$, with

\be
W^{\pm}_M=\frac{1}{\sqrt{2}}(W^1_M\mp i W^2_M).
\ee

\noindent This identiy are useful 

\be
\sigma^aW^a_M= W_{M}^3T_3 + \sqrt{2}(W^+_{M}T^++W^-_MT^-).
\ee

\noindent   $T_3$ and $T^{\pm}$ are separately,

\[
T^3 = \left(
\begin{array}{cc}
1 & 0 \\
0 & -1 
\end{array}  \right)
\ \ \ 
T^+ = \left(
\begin{array}{cc}
0 & 1 \\
0 & 0 
\end{array}  \right)
\ \ \ 
T^- = \left(
\begin{array}{cc}
0 & 0 \\
1 & 0 
\end{array}  \right)
\]

\noindent Also define $V_{\text{CKM}}$ and $V_{\text{PMNS}}$ as

\be
V_{\text{CKM}} = U_u^\dag U_d  \  \ \  \ \ \ \ \ \  V_{\text{PMNS}} = U_\nu^\dag U_l
\ee

\noindent as well as 

\be
W^1_MW_N^2-W^2_MW^1_N &=& i( -W^+_MW_N^- + W^-_MW_N^+) \n \\
W^1_MW_N^1+W^2_MW^2_N &=& W^+_MW_N^- + W^-_MW_N^+
\ee

Writing all Lagrangian terms in the phyiscal basis. The Lagrangian terms giving rise to vertices  W-W-W and W-W-W-W are

\be
{\mathcal L}_{W^3_M}&=& -ig(\partial_\mu W_N^3W^+_\rho W^-_K)g^{\mu\rho}g^{\prime NK} + \text{cyclic in (3, +, -)} \n \\
{\mathcal L}_{W^4_M}&=& \frac{g^2}{2}\left[g^{\mu\nu}g^{\rho\sigma}W^+_{\mu}W^+_{\nu}W^-_\rho W^-_\sigma -g^{\mu\nu}g^{P\Sigma}W^+_{\mu}W^-_{\nu}W^+_PW^-_\Sigma \right] -\frac{\lambda_h}{4} (\phi^+\phi^-)^2  \n \\
&&-\frac{g^2}{2}\left[g^{\mu\nu}g^{\prime P\Sigma} W^+_{\mu}W^-_{\nu}W^3_PW^3_\Sigma-g^{\mu\nu}g^{P\Sigma}W^+_{\mu}W^3_{\nu}W^-_PW^3_\Sigma + (\mu\leftrightarrow P, \nu \leftrightarrow \Sigma)\right] \n \\
&&-\frac{\lambda_h}{4}\phi^+\phi^-(\phi^3)^2  \\ 
&&+\frac{g^2}{8}W_{\mu}^3W_{\nu}^3\phi^3\phi^3g^{\mu\nu}-\frac{\lambda_h}{16}(\phi^3)^4 \n 
\ee

\noindent  The Lagrangian terms giving rise to vertices  h-W-W and h-h-W-W are

\be
{\mathcal L}_{hW_M^2}&=& \frac{g}{2}(\partial_{\mu}hW_{\nu}^3\phi^3-\partial_{\mu}\phi^3W_{\nu} h )g^{\mu\nu} +\frac{1}{2}gm_WhW_{\mu}^3W^{\mu 3}-\frac{\lambda_h}{4} vh (\phi^3)^2 \n \\
&&\left[\frac{g}{2}g^{\mu\nu}(\partial_{\mu}hW_{\nu}^+\phi^- -\partial_{\mu}\phi^-W_{\nu}^+ h)+(+\leftrightarrow -) \right] + gm_Wh W_{\mu}^+W^{\mu -} -\frac{1}{2}\lambda_h vh \phi^+\phi^- \\
{\mathcal L}_{h^2W_M^2}&=& \frac{g^2}{8}g^{\mu\nu}h^2(2W_{\mu}^+W_{\nu}^-+W_{\mu}^3W_{\nu}^3)-\frac{\lambda_h}{8}h^2(2\phi^+\phi^-+(\phi^3)^2)\n
\ee

\noindent The Lagrangian terms for the quark sector become 

\be
{\mathcal L}_{ffW_M}&=& -\frac{g}{2}(\bar{u}_L\slashed{W}^3u_L-\bar{d}_L\slashed{W}^3d_L)+\left[\frac{i}{\sqrt{2}}(y_u\bar{u}_Lu_R\phi^3 -y_d\bar{d}_Ld_R\phi^3)+\text{h.c.}\right] \n\\
&&-\frac{g}{\sqrt{2}}(\bar{u}_L\slashed{W}^+V_{\text{CKM}}d_L+\text{h.c.})+\left[\frac{i}{\sqrt{2}}(y_u\bar{u}_LV_{\text{CKM}}d_R\phi^+ +y_d\bar{d}_LV^\dag_{\text{CKM}}u_R\phi^-)+\text{h.c.}\right] \\
{\mathcal L}_{ffh}&=& -\frac{1}{\sqrt{2}} \left(y_u\bar{u}_Lu_R h+ y_d\bar{d}_Ld_Rh \right) + \text{h.c.}\n
\ee

\noindent For the lepton sector we simply make the replacement $(u\rightarrow \nu, d \rightarrow l)$ and $V_{\text{CKM}}\rightarrow V_{\text{PMNS}}$.

Finally we comment on a subtlety in the derivation of Feynman rules: since W and $\phi$ are simply different components of the same physical fields, it's necessary to sum over all possible contractions in deriving Feynman rules.   This operation gives an additional overal factor to the Feynman rules. For example, the overal factor of vertex $hhhh$ is $4!=24$. However, sometimes it  requires writing different terms explicitly.  This is especially true in the case of gauge-goldstone fields, in which the gauge components and goldstone components appear to be different fields.    For example, in ${\mathcal L}_{hW_M^2}$ of  Eq.(\ref{eq:vertices_20}), the term   $\frac{g}{2}(\partial_{\mu}hW^a_{\nu}\phi_ag^{\mu\nu}-\partial_{\mu}\phi^aW_{\nu}^ah g^{\mu\nu})$ gives rise to an overal factor $2$, which is taken into account by adding a different term with $\phi^a \leftrightarrow W^a_{\mu}$ in the Feynman rule. While for $W^{\pm}_M$  they have been written out explicitly, for $W^3_M$ we still need to take into account  the term with $\phi^3 \leftrightarrow W^3_{\mu}$.

\

\

 \section{ On-shell Match for 3-Point Amplitudes }
 \label{sec:3-point}
 
 \subsection{On-shell Gauge Symmetry}
 \label{sec:gauge_sym}
 


Having derived all the Feynman rules in a physical gauge.  We proceed to analyse the on-shell gauge symmetry, especially how they are reflected in 3-point amplitudes.  So far we have concluded that  a longitudinal polarization in a physical gauge is composed of gauge components $\epsilon_n^{\mu}$ and goldstone component $i/-i$, as in Eq.(\ref{eq:pol_1}).  We choose $n^2=0$, i.e. light-cone gauge,    the longitudinal polarization vector in 5-component becomes:  

\be\label{eq:long_pol_1}
\epsilon_{1L}^{M}= \left(
\begin{array}{c}
\epsilon_{n_1}^{\mu} \\
i 
\end{array}
\right)
\ee

\noindent for incoming particles, and $\epsilon_L^{*M}$ for outgoing particles.  Here $\epsilon^{\mu}_{n_1}=-\frac{m_W}{n_1\cdot k}n_1^{\mu}$.  On the other hand, we already have the standard form for the longitudinal polarization vector,  which can be written in the 5-component format  as 

\be\label{eq:long_pol_2}
\epsilon_{2L}^{M}= \left(
\begin{array}{c}
\epsilon_{n_2}^{\mu} +\frac{k^{\mu}}{m_W}\\
0
\end{array}
\right)  
\ee

\noindent with the 5th component to be $0$ and $n_2^{\mu}=(1,-\frac{\vec{k}}{|\vec{k}|})$.  If we choose $n_1$ in Eq.~(\ref{eq:long_pol_1}) as $n_1=n_2$,  $\epsilon_{1L}^M$ and $\epsilon_{2L}^M$ are related with each other by

\be\label{eq:gauge_equiv}
\epsilon_{2L}^M= \epsilon^M_{1L} - \frac{k^M}{m_W}
\ee

Meanwhile since S-matrix is gauge-invariant,  the two forms of longitudinal polarizations have to give the same S-matrix, i.e.,

\be\label{eq:gauge_sym_1}
\epsilon_{2L}^{M_1}...\epsilon_{2L}^{M_i}S_{M_1...M_i}(k_1...k_i...) = \epsilon_{1L}^{M_1}...\epsilon_{1L}^{M_i}S_{M_1...M_i}(k_1...k_i...)
\ee

\noindent Plugging in Eq.(\ref{eq:gauge_equiv}),  Eq.(\ref{eq:gauge_sym_1}) is equivalent to 

\be\label{eq:gauge_sym_2}
k_1^{M_1}...k_i^{M_i}S_{M_1...M_i}(k_1...k_i...)=0.
\ee

\noindent Interestingly, Eq.(\ref{eq:gauge_sym_2}) share similar form as the on-shell gauge symmetry for massless gauge theory, with $M\rightarrow \mu$.   Thus, we can appropriately call it ``on-shell gauge symmetry" for massive gauge theory. 

Ignoring the terms of $O(\frac{m_W}{n\cdot k })$, Eq.(\ref{eq:gauge_sym_1}) reduced to 

\be\label{eq:GET_3}
\epsilon_{2L}^{M_1}...\epsilon_{2L}^{M_i}S_{M_1...M_i}(k_1...k_i...) = S(\phi_1...\phi_i...) + O(\frac{m_W}{n\cdot k })
\ee

\noindent So on-shell gauge symmetry reduces to goldstone equivalence theorem as in  Eq.(\ref{eq:GET}).

In this paper we don't intend to give a complete proof of on-shell gauge symmetry Rather,  we only set to prove  that Eq.~(\ref{eq:gauge_sym_2})  is satisfied for all on-shell 3-point amplitudes: $WWW$, $hWW$ and $ff'W$.  

The condition of ``on-shell" needs some extra comments, as it's not always kinematically possible to put all particles on-shell for 3-point amplitudes.  This constraint is usually  bypassed by analytical continuation of making momenta complex. Nevertheless,  for $hWW$ and $ff'W$, we can also put amplitudes on-shell through the  analytical continuation of  parameters in the theory.  For example, for a decay process $h\rightarrow W^+W^-$, the Higgs mass has to satisfy the on-shell condition $m_h\geq 2m_W$, with $m_h\sim \lambda_h v$, $m_W\sim gv$.  We can think of the amplitude as the function of parameters of theory: $\mathcal{M}=\mathcal{M}(\lambda_h, v, g)$. We can first choose the parameters to satisfy the on-shell condition, but the resulting amplitude will have to be the same for all possible values of parameters within the perturbative limits.   The same argument can be applied to $f\rightarrow f' W$,  whose amplitude can be seen as the function of $y_f, y_{f'}, g, v$: $\mathcal{M}=\mathcal{M}(\lambda_f, \lambda_f', g, v)$.  Notice this  argument doesn't apply to $WWW$,  since the  amplitude is controlled by only one coupling $g$.  It's not possible to  adjust $g$ to put the all the particles on-shell.  

For the convention,   we absorb the intrinsic minus sign between gauge components and goldstone components in amplitudes to the definition of polarization vectors,  which are given by Eq.(\ref{eq:pol_2}) with $g^{44}=1$, $n^2=0$,

\be
 g^{44}=1:   \  \  \    \    \   \   \ 
\epsilon^M_L=\left(\begin{array}{c}
\epsilon_n^{\mu}\\
 -i 
 \end{array}\right)  
 \  \   \ 
\epsilon^{*M}_L=\left(\begin{array}{c}
\epsilon_n^{\mu} \\
 i 
 \end{array}\right)  \n
\ee

\noindent with $\epsilon_n^{\mu}=-\frac{m_W}{n\cdot k }n^{\mu}$ satisfying $k\cdot \epsilon_n=-m_W$.

  In proving the on-shell gauge symmetry for 3-point amplitudes,  $k^M/k^{*M}$ also need to be redefined to

\[ 
k_{L}^{M}= \left(
\begin{array}{c}
k^{\mu} \\
i m_W
\end{array}
\right)  \  \   \ \  \  \  \  \   k_{L}^{*M}= \left(
\begin{array}{c}
k^{\mu} \\
-i m_W
\end{array}
\right)
\]

 However,  in the transverse condition and on-shell condition for $k^M$, 
the intrinsic minus sign between gauge components and goldstone component still need to be taken into account,  i.e. we have 

\be
k^M\epsilon_M^{*L}&=&k^{*}_M\epsilon_M^L=k\cdot \epsilon_n + m_W =0  \n  \\ 
k^M k_M^{*}&=&k^{*M}k_M=k^2-m_W^2=0    \n
\ee

We also extract the factor ``i" out for S-matrix by defining $S=i\M$.  Our convention is that  all particles are incoming. 

\

\

\noindent{\bf $\textbf{h-W-W}$}

The on-shell gauge symmetry for 3-point amplitudes  can be written as

\be\label{eq:3-point}
i\mathcal{M}(1^{s_1}2^{s_2}3^{s_3})|_{\epsilon_{s_i}^M\rightarrow \frac{k_i^M}{m_W} } = 0
\ee 

$i$ denotes any particles being $W$ bosons.  To prove it,  we start with $\mathcal{M}(hWW)$, with only one $W$ replaced by $\frac{k^M}{m_W}$.   With one particle being the Higgs and another particle  being $\frac{k^M}{m_W}$,  there are two cases for the polarizations of particle 2: a) transverse b) longitudinal.   We start with case a) with $s_2=\pm$.  In this case  $\epsilon_{s_2}^4=0$, so there is no goldstone component contribution from particle $2$, but $\frac{k_3^4}{m_W}=i$.  

\be
i\mathcal{M}(1^{h}2^{s_2=\pm}3^{s_3})|_{\epsilon_{s_3}^M\rightarrow \frac{k_3^M}{m_W} } &=&igm_W\epsilon_2^{\pm}\cdot \frac{k_3}{m_W} +  \frac{g}{2}((k_1-k_3)\cdot \epsilon_2^{\pm}) (i) \n \\
&=& ig(k_3\cdot \epsilon_2^{\pm}+ \frac{1}{2}(-k_2-2k_3)\cdot \epsilon_2^{\pm} ) \n \\
&=&0 \n
\ee

In the second step we used energy-momentum conservation $k_1+k_2+k_3=0$, in the third step we used the transverse condition for particle 2: $k_2\cdot \epsilon_2^{\pm}=0$. 

Then we turn to case b)  $s_2=L$, so the polarization vector has both gauge components $\epsilon_{2n}^{\mu}$ and goldstone component $\epsilon_2^4=-i$, the amplitude is 

\be
i\mathcal{M}(1^{h}2^{s_2=L}3^{s_3})|_{\epsilon_{s_3}^M\rightarrow \frac{k_3^M}{m_W} } &= &   igm_W\epsilon_2^n\cdot\frac{k_3}{m_W} +
\frac{g}{2}((k_1-k_3)\cdot \epsilon_2) (i) + \frac{g}{2}(k_1-k_2)\cdot \frac{k_3}{m_W} (-i) -ig\frac{m_h^2}{2m_W}i\cdot(-i) \n \\
&=& ig (\epsilon_2^n\cdot k_3-\epsilon_2^n\cdot k_3 -\frac{1}{2} k_2\cdot \epsilon_2^n -\frac{(k_1-k_2)(k_1+k_2)}{2m_W}-\frac{m_h^2}{2m_W}) \n 
\ee

To further simplify, we need first to  make use of on-shell condition for $k_1$ and $k_2$:  

\[ (k_1-k_2)(k_1+k_2)=k_1^2-k_2^2=m_h^2-m_W^2 \]

as well as the transverse condition for $\epsilon^M_{2L}$:  $k_2\cdot \epsilon_2^n=-m_W$, which is another expression of $k^{*M}\cdot \epsilon_L^M=0$. Plugging in, the amplitude becomes

\be
i\mathcal{M}(1^{h}2^{s_2=L}3^{s_3})|_{\epsilon_{s_3}^M\rightarrow \frac{k_3^M}{m_W} }  &=& ig(\frac{m_W}{2}+\frac{m_h^2-m_W^2}{2m_W}-\frac{m_h^2}{2m_W}) \n \\
&=&0, \n
\ee

\

\

\noindent{\bf $\textbf{$f$-$f'$-W}$}

We then proceed to prove the on-shell gauge symmetry Eq.~(\ref{eq:3-point}) for the amplitude of $ffW$,

\be
i\mathcal{M}(1^{s_1}2^{s_2}3^{s_3})|_{\epsilon_{s_3}^M\rightarrow \frac{k_3^M}{m_W} } = 0 \n
\ee

with particle 1 and 2 being fermions,  particle 3 being W boson.  Since only particle 3 is W boson,  the identity is relatively easy to prove.  Writing the amplitude with gauge components and the amplitude with the goldstone component separately,  Eq.~\ref{eq:3-point} becomes, 

\be\label{eq:3-point-2}
i\mathcal{M}(1^{s_1}2^{s_2}3^{s_3})|_{\epsilon_{s_3}^M\rightarrow \frac{k_3^M}{m_W} }  = \mathcal{M}(1^{s_1}2^{s_2}3^{s_3})|_{\epsilon_{s_3}^{\mu}\rightarrow \frac{k_3^{\mu}}{m_W}, \epsilon_{s_3}^4\rightarrow 0 } + \mathcal{M}(1^{s_1}2^{s_2}3^{s_3})|_{\epsilon_{s_3}^{\mu}\rightarrow 0, \epsilon_{s_3}^4\rightarrow -i }=0
\ee

\noindent The first term is the amplitude from the fermion-fermion-gauge vertex, the second term is from the fermion-fermion-goldstone vertex.    
Let's check Eq.~(\ref{eq:3-point-2}) explicitly.  First look at the first term,  to fix all the momenta to be incoming, we choose particle 1 to be a d-type anti-quark, particle 2 to be a d-type quark, then the W boson has to be $W^+$, we also set the CKM matrix to be $\bf 1$,   the amplitude becomes 

\be\label{eq:3-point-3}
i\M|_{\epsilon_{s_3}^{\mu}\rightarrow \frac{k_3^{\mu}}{m_W}, \epsilon_{s_3}^4\rightarrow 0 } &=& -i\frac{g}{\sqrt{2}}\bar{\nu}_L^{s_1}\gamma^{\mu}u_L^{s_2}\frac{k_{3\mu}}{m_W}\n \\
   &=& i\frac{g}{\sqrt{2}} \bar{\nu}_L^{s_1}(\slashed{k}_1+\slashed{k}_2)u_L^{s_2}/m_W \n \\
   &=& i\frac{g}{\sqrt{2}}(-\frac{m_d}{m_W}\bar{\nu}_R^{s_1}u_L^{s_2}+\frac{m_u}{m_W}\bar{\nu}^{s_1}_Lu_R^{s_2} )
\ee

\noindent In the second equality, we made use of the energy-momentum conservation $k_1+k_2+k_3=0$, in the third equality we made use of equation of motion for fermions: $\slashed{k}_2u_{L/R}(k_2)=m_uu_{R/L}(k_2)$ as well as $\bar{\nu}_{L/R}(k_1)\slashed{k}_1=-\bar{\nu}_{R/L}(k_1)m_d$.  We then look at  the second term in Eq.~(\ref{eq:3-point-2}), with  $\epsilon_3^4=-i$, we have

\be\label{eq:3-point-4}
i\M|_{\epsilon_{s_3}^{\mu}\rightarrow 0, \epsilon_{s_3}^4\rightarrow -i }&=& -1\cdot (-i)\frac{g}{\sqrt{2}}(\frac{m_d}{m_W}\bar{\nu}_R^{s_1}u_L^{s_2}-\frac{m_u}{m_W}\bar{\nu}^{s_1}_Lu_R^{s_2} )  \n \\
&=& -i\frac{g}{\sqrt{2}}(-\frac{m_d}{m_W}\bar{\nu}_R^{s_1}u_L^{s_2}+\frac{m_u}{m_W}\bar{\nu}^{s_1}_Lu_R^{s_2} )
\ee

\noindent Here we made use of $y_f=\frac{m_f}{\sqrt{2}m_W}$. Combining Eq.~(\ref{eq:3-point-3}) and Eq.~(\ref{eq:3-point-4}), we finish the proof of  Eq.~(\ref{eq:3-point-2}).   Although we only went through the example of  $\bar{d}_1u_2W^+_3$, it can be checked straightforwardly that Eq.~(\ref{eq:3-point-2}) is satisfied for all other cases. Having ignoring the CKM matrix $\bar{u}_1d_2W^-_3$ is identical to $\bar{d}_1u_2W^+_3$.  For neutral current, i.e. W being $W^3$, 
the proof is also identical except we need to replace $\gamma^{\mu}$ with $\frac{1}{\sqrt{2}}\gamma^{\mu}T_3$ for the  fermion-fermion-gauge vertex, and replace $\gamma^5$ with $\frac{1}{\sqrt{2}}\gamma^5T_3$ for the fermion-fermion-goldstone vertex.  It's also not hard to see that the conclusion also applies to the SM with the gauge group being $SU(2)_L\times U(1)_Y$, in which case the only difference is the  neutral current case.  For $ff\gamma$, $i\M|_{\epsilon_3^{\mu}\rightarrow k^{\mu}}=0 $ according to ward identity.  For $ffZ$, the fermion-fermion-goldstone vertex is identical to the $SU(2)$ case; for the  fermion-fermion-gauge vertex, ignoring the overall factor difference,  there is an additional term of vector current $Q_f\sin^2\theta_W$ relative to the $SU(2)$ case. Nevertheless, this term gives 0 when $k^{\mu}$ dots into the S-matrix because the interaction is vector-like.  So we conclude Eq.~(\ref{eq:3-point-2}) is satisfied for the SM too. Indeed, since the argument is very general, we  expect Eq.~(\ref{eq:3-point-2}) applies to any massive gauge theory with Higgs mechanism.

\

\

 \noindent{\bf $\textbf{W-W-W}$}
 
 Next we proceed to prove Eq.~(\ref{eq:3-point-2}) for the amplitude of $WWW$. Stripping of the overall factor of $-ig$,  the general amplitude of $WWW$ can be written as

  \be 
 i\mathcal{M}(1^{s_1}2^{s_2}3^{s_3})= (\epsilon_1\cdot \epsilon_2-\frac{1}{2}\epsilon_1^4\cdot \epsilon_2^4)[(p_1-p_2)\cdot \epsilon_3] +
 \text{cyclic}
  \ee

  Replacing one of the polarizations are replaced by $\frac{k^M}{m_W}$,  there are three different cases for the other two polarizations:  a) two transverse; b) one transverse and one longitudinal;  c) two longitudinal.   We start from case a), since both particle 1 and particle 2 are transverse their goldstone components are $0$: $\epsilon_1^4=\epsilon_2^4=0$.  Consequently, there is no goldstone contribution in this case,  so we have

\be
i\mathcal{M}(1^{s_1=T}2^{s_2=T}3^{s_3})|_{\epsilon_{s_3}^M\rightarrow \frac{k_3^M}{m_W} }= i\mathcal{M}(1^{s_1=T}2^{s_2=T}3^{s_3})|_{\epsilon_{s_3}^{\mu}\rightarrow \frac{k_3^\mu}{m_W}, \epsilon_{s_3}^4 \rightarrow 0 }
\ee

This means the on-shell gauge symmetry for $1^T2^T3^{s_3}$ is directly analogue to the massless case, with only gauge vertex contributing. 

\be
i\mathcal{M}(1^{s_1=T}2^{s_2=T}3^{s_3})|_{\epsilon_{s_3}^{\mu}\rightarrow k_3^\mu, \epsilon_{s_3}^4 \rightarrow 0 }
&=& (\epsilon^T_1\cdot \epsilon_2^T) [(k_1-k_2)\cdot (-k_1-k_2)] + \epsilon_2^T\cdot k_3 [(k_2-k_3)\cdot \epsilon_2^T]  \n \\
&&+ k_3\cdot \epsilon_1^T[(k_3-k_1)\epsilon_2^T]  \n \\
&=& 0 - 2\epsilon_2^T\cdot k_3 \  k_3\cdot \epsilon_1^T + 2 k_3\cdot \epsilon_1^T \  k_3\cdot \epsilon_2^T \n \\
&=& 0 
\ee

In the second step we made use of energy-momentum conservation,  on-shell conditions and transverse conditions for particle 1 and 2 respectively, 

\be
&&k_1+k_2+k_3=0 \n \\
&&k_1^2-k_2^2=m_W^2-m_W^2=0 \\
&&k_1\cdot  \epsilon_1^T =  k_2\cdot \epsilon_2^T =0\n 
\ee

Next we turn to case b) with particle 1 being transverse and particle 2 being longitudinal,  we get 

\be
i\mathcal{M}(1^{s_1=T}2^{s_2=L}3^{s_3})|_{\epsilon_{s_3}^M\rightarrow \frac{k_3^M}{m_W} } &=& \epsilon_1^T\cdot \epsilon_2^n \ \frac{(k_1-k_2)\cdot k^3}{m_W} +( \epsilon_2^n\cdot \frac{k_3}{m_W}-\frac{1}{2})\  (k_2-k_3)\cdot \epsilon_1^T 
+\frac{k_3\cdot \epsilon_1^T}{m_W}\ (k_3-k_1)\cdot \epsilon_2^n  \n \\
&=& 0+ \frac{\epsilon_2^n\cdot k_3}{m_W} \ (-k_1-2k_3)\cdot \epsilon_1^T -\frac{1}{2}(-k_1-2k_3)\cdot \epsilon_1^T + \frac{k_3\cdot \epsilon_1^T}{m_W} \ (2k_3+k_2)\cdot \epsilon_2^n \n \\
&=& \frac{\epsilon_2^n\cdot k_3 \  2k_3\cdot \epsilon_1^T}{m_W} + \frac{-k_3\cdot \epsilon_1^T\  2k_3\cdot \epsilon_2^n}{m_W}-k_3\cdot \epsilon_1^T + k_3\cdot \epsilon_1^T \n \\
&=& 0
\ee

Again we have used the conditions of all the particles being on-shell.  For the longitudinal state $\epsilon_{2L}^M$  the on-shell condition 
implies $k_2\epsilon_{2n}=-m_W$. 

Finally,  we turn to case c) with both particle 1 and 2 being longitudinal polarizations,  we have 

\be
i\mathcal{M}(1^{s_1=L}2^{s_2=L}3^{s_3})|_{\epsilon_{s_3}^M\rightarrow \frac{k_3^M}{m_W} } &=& (\epsilon_1^n\cdot \epsilon_2^n+\frac{1}{2}) \ \frac{(k_1-k_2)\cdot k^3}{m_W} +( \epsilon_2^n\cdot \frac{k_3}{m_W} -\frac{1}{2}(-i)\cdot i)\  (k_2-k_3)\cdot \epsilon_1^n  \n \\
&+&(\frac{k_3\cdot \epsilon_1^n}{m_W}-\frac{1}{2}(-i)\cdot i)\ (k_3-k_1)\cdot \epsilon_2^n  \n \\
&=& 0+ \frac{\epsilon_2^n\cdot k_3}{m_W} \ (-k_1-2k_3)\cdot \epsilon_1^n -\frac{1}{2} (-k_1-2k_3)\cdot\epsilon_1^n  \n \\
&&+ \frac{k_3\cdot \epsilon_1^n}{m_W} \ (2k_3+k_2)\cdot \epsilon_2^n  -\frac{1}{2} (2k_3+k_2)\cdot\epsilon_2^n \n \\
&=& \frac{k_3\cdot \epsilon_2^n\ (-2)k_3\cdot \epsilon_1^n}{m_W} + k_3\cdot \epsilon_2^n+-\frac{1}{2}m_W + k_3\cdot \epsilon_1^n \n \\
&&+\frac{k_3\cdot \epsilon_1^n\  2k_3\cdot \epsilon_2^n}{m_W} -k_3\cdot \epsilon_1^n +\frac{1}{2}m_W -k_3\cdot \epsilon_2^n \n \\
&=& 0
\ee

Thus combining case a), case b) and case c), we  have proved that on-shell gauge symmetry is satisfied for all possible 3-point amplitudes:  $h$-$W$-$W$, $f$-$f'$-$W$, $W$-$W$-$W$.  However, our proof  still has two loop holes: the first one is only one W state is replaced by $\frac{k^M}{m_W}$, the second one is particles are assumed to be incoming.  Here we demonstrate neither of the two assumptions affect the conclusion.  Starting with the first one,  to prove the general case of arbitrary number of polarizations of W being replaced by $\frac{k^M}{m_W}$, we need only notice that  by replacing the gauge components $\epsilon_n^{\mu}$ with $k^\mu$, the transverse condition for the longitudinal polarization 

\[k^{*M}\epsilon_{LM}=0\]

\noindent turns to the on-shell condition for $k^M$, 
 
 \[k^{*M}k_{M}=0.\]
 
\noindent Since in the proof above, the only conditions we used are   on-shell condition for $k^M$ and transverse conditions for $\epsilon^{M}_{s}$, the proof  of Eq.~(\ref{eq:3-point}) is exactly the same for multiple polarization vectors being replaced by $\frac{k^M}{m_W}$.  

The second loophole is  automatically fixed if crossing symmetry is satisfied. Under $k\rightarrow -k$, the incoming longitudinal state becomes $\epsilon_L^M(-k)= -\epsilon_L^{*M}(k)$, as $\epsilon_n^{\mu}(-k)=-\epsilon^{\mu}_n(k)$. So we get the  longitudinal polarization vector for the outgoing state up to a minus sign.  Energy-momentum conservation becomes 

\[\sum_ik_i+k=0\rightarrow \sum_i k_i+(- k)=0 \] 

\noindent Thus we obtain the amplitude for one particle being outgoing if its momentum is under $k\rightarrow -k$.   So crossing symmetry is indeed satisfied. Moreover, $k^{M}/k^{*M}$ turns to $-k^{*M}/-k^{M}$ under $k\rightarrow -k$. Therefore, we finished our proof of on-shell gauge symmetry for all 3-point amplitudes. 





\

\

\

\subsection{$1\rightarrow 2$ Splitting Amplitudes}
\label{sec:split_amp}










In this section we are demonstrating how to use the new Feynman rules and the on-shell match condition from on-shell gauge symmetry to do calculations.  Our examples are $1\rightarrow 2 $ collinear splitting amplitudes involving longitudinal vector bosons: $W_L\rightarrow W_L W_L$,  $h\rightarrow W_L W_L$ and $f\rightarrow f' W_L$.  Those splitting amplitudes have been calculated in \cite{Chen:2016wkt}. However,  as we will see, with the new prescriptions the calculations become   largely simplified. 

When external states of a process become collinear with each other,  one internal lines of one of the Feynman diagrams will approach its pole or mass singularity.  The amplitude can then be factorised in the following way 

\be
i{\mathcal M}=  \sum_{s} i{\mathcal M}_{split}^s \cdot \frac{i}{k_3^2-m_3^2}\cdot  i{\mathcal M}_0^s +\text{power suppressed}
 \ee
 
Thus  the splitting amplitude ${\mathcal M}_{\text{split}}$ should be evaluated on-shell.  The collinear splitting amplitudes ${\mathcal M}_{split}$ are related to  collinear splitting functions in the following way\cite{Chen:2016wkt},

 \be\label{eq:split_fun}
\frac{d{\mathcal P} }{dzdk_T^2} \propto |{\mathcal M}_{split}|^2
\ee

So evaluating collinear splitting functions is reduced to evaluating splitting amplitudes.

\noindent {\large \bm{$h\rightarrow W^+_LW^-_L$}}

The splitting amplitude for $h(k_3)\rightarrow W^+_L(k_1)W^-_L(k_2)$ can be more conveniently calculated using the polarization vector $\epsilon_L^{\mu}=\frac{k^{\mu}}{m_W}-\frac{m_W}{n\cdot k}n^{\mu}$,  and evaluating the splitting amplitude by treating all particles ``on-shell".

\be
i{\mathcal M}(h\rightarrow W^+_LW^-_L)&=&igm_W\left(\frac{k_2^{\mu}}{m_W}+\epsilon_{2n}^{\mu}\right)\left(\frac{k_{1\mu}}{m_W}+\epsilon_{n_1 \mu}\right) \n \\
                               &=&  igm_W\left(\frac{k_3^2-k_2^2-k_1^2}{2m_W^2}+\frac{k_2\cdot \epsilon_{n_1}+k_1\cdot \epsilon_{n_2}}{m_W}+\epsilon_{n_2}\epsilon_{n_1}\right) \n \\
                               &\overset{\text{onshell}}{=}& igm_W\left(\frac{m_h^2-2m_W^2}{2m_W^2}-\frac{k_2\cdot n_1}{k_1\cdot n_2}-\frac{k_1\cdot n_2}{k_2\cdot n_1}+m_W^2\frac{n_2\cdot n_1}{(n_2\cdot k_2)(n_1\cdot k_1)}\right)\n \\
                               &=& igm_W\left(\frac{m_h^2-2m_W^2}{2m_W^2}-\frac{\zb}{z}-\frac{z}{\zb}\right)
\ee

In the third step we made use of on-shell conditions  $k_3^2=m_h^2$, $k_1^2=m_W^2$ and $k_2^2=m_W^2$; in the final step we choose $n_1=n_2=n_3=n$, and  define energy fraction of $k_1/k_3$ to $k_2/k_3$ as 

\be\label{eq:z/zb}
z=\frac{n\cdot k_1}{n\cdot k_3}\  \  \  \  \  \  \  \  \  \    \zb=\frac{n\cdot k_2}{n\cdot k_3}
\ee

In the limit of particles 1, 2 and 3 are massless, as well as $k_1$, $k_2$, $k_3$ are collinear with each other, we have $\zb=1-z$.

After reorganization, we have 

\be
i{\mathcal M}(h\rightarrow W^+_LW^-_L)=igm_W\frac{1}{z\zb}\left(\frac{m_h^2}{2m_W^2}z\zb-(1-z\zb)\right)
\ee

\

\noindent {\large \bm{$f\rightarrow f' W^+_L$} }

Similar to $h\rightarrow W^+_LW^-_L$, the splitting amplitude can be evaluated ``on-shell" using the polarization vector $\epsilon_L^{\mu}=\frac{k^{\mu}}{m_W}-\frac{m_W}{n\cdot k}n^{\mu}$.  For the interaction between fermion current and gauge boson is given by   ${\mathcal L}=\frac{g}{\sqrt{2}}\bar{\psi}_1\gamma^{\mu}P_L\psi_2W_{\mu}$, we have the splitting amplitude to be

\be\label{eq:ffWL}
i{\mathcal M}(f^{s_3}\rightarrow f'^{s_2} W^+_L)&=&i\frac{g}{\sqrt{2}}\bar{u}^{s_2}_L(k_2)\gamma^{\mu}u^{s_3}_L(k_3)
\cdot \left(\frac{k_{1\mu}}{m_W}+\epsilon_{n_1\mu}\right) \n \\
     &=&i  \frac{g}{\sqrt{2}m_W}\bar{u}^{s_2}_L(k_2)(\slashed{k}_3-\slashed{k}_2)u^{s_1}_L(k_3)
     -i\frac{gm_W}{\sqrt{2}n_1\cdot k_1}  \bar{u}_L(k_2)\slashed{n}_1u_L(k_3) \n \\
    &\overset{\text{onshell}}{=}& i\frac{g}{\sqrt{2}m_W}(m_2\bar{u}^{s_2}_R(k_2)u_L^{s_3}(k_3)-m_3\bar{u}^{s_2}_L(k_2)u^{s_3}_R(k_3)) \n \\
    && -i\frac{g}{\sqrt{2}}\frac{m_W}{n_1\cdot k_1}\bar{u}_L^{s_2}(k_2)\slashed{n}_1u^{s_3}_L(k_3)
\ee

In the second line, we made use of equations of motion for the fermions. The first two terms give the contribution of the goldstone component, as can be seen by the factor $\frac{gm_{f_i}}{\sqrt{2}m_W}=y_{f_i}$, with $i=2,3$.  To continue the calculation, we need the explicit form of the fermion wave function:

\be
u_L^{-\frac{1}{2}}(k) =u^{\frac{1}{2}}_R (k) = \sqrt{n\cdot k} \xi  &&\hspace{2cm}  u^{-\frac{1}{2}}_R(k) =u^{\frac{1}{2}}_L (k) = \frac{m}{\sqrt{n\cdot k}}\xi  \n \\
\bar{u}^{-\frac{1}{2}}_L(k) =\bar{u}^{\frac{1}{2}}_R (k) = \sqrt{n\cdot k} \xi^\dag &&\hspace{2cm}  \bar{u}^{-\frac{1}{2}}_R(k) =\bar{u}^{\frac{1}{2}}_L (k) = \frac{m}{\sqrt{n\cdot k}}\xi^\dag  
\ee

\noindent Here $n^{\mu}=(1,-\frac{|\vec{k}|}{\vec{k}})$. 

We take $s_1=s_2=-\frac{1}{2}$ as the example,  at the collinear limit $k_1\simeq k_2\simeq k_3$,  we have $n_1\simeq n_2 \simeq n_3=n=(1,0,0,-1)$ with z direction along $\vec{k}_3$, and $\xi^\dag_2\xi_3\simeq \xi^\dag \xi=1$

\be\label{eq:ffW_2}
\bar{u}^{-\frac{1}{2}}_R(k_2)u_L^{-\frac{1}{2}}(k_3)&=& m_2\sqrt{\frac{n\cdot k_3}{n\cdot k_2}} \xi^\dag_2 \xi_3 = m_2\frac{1}{\sqrt{\zb}} \n \\
  \bar{u}^{-\frac{1}{2}}_L(k_2)u_R^{-\frac{1}{2}}(k_3)&=& m_1\sqrt{\frac{n\cdot k_2}{n\cdot k_3}}\xi^\dag_2 \xi_3= m_3\sqrt{\zb}
\ee

Here we have used the definition of $z/\zb$ as in Eq.~({\ref{eq:z/zb}}). We also need,

\be\label{eq:ffW_3}
\bar{u}_L^{-\frac{1}{2}}(k_2)\slashed{\epsilon}_{n_1}u^{-\frac{1}{2}}_L(k_3) =  -\frac{m_1}{n\cdot k_1}\sqrt{(n\cdot k_3)(n\cdot k_2)} \xi^\dag n\cdot \sigma\xi = -2m_1\frac{\sqrt{\zb}}{z} 
\ee

Here we have used 

\[  \xi^\dag_{-\frac{1}{2}} n\cdot \sigma\xi_{-\frac{1}{2}} =\xi^\dag (1-(-1))\xi  = 2  \]

 Plug  Eq.~(\ref{eq:ffW_2}) and Eq.~(\ref{eq:ffW_3}) into Eq.~(\ref{eq:ffWL}), we get, 

\be\label{eq:ffW1}
i{\mathcal M}(f^{-\frac{1}{2}}\rightarrow f'^{-\frac{1}{2}} W^+_L) 
&=& \frac{ig}{\sqrt{2}}\frac{1}{\sqrt{\zb}z} (\frac{m_2^2}{m_W}z- \frac{m_3^2}{m_W} z\zb-2 m_W\zb) \n \\
 &=& i (y_{f_2}m_2\frac{1}{\sqrt{\zb}}- y_{f_1}m_1\sqrt{\zb}- \frac{g}{\sqrt{2}} 2m_W\frac{\sqrt{\zb}}{z}) 
\ee

Similarly,  for $s_2=s_3=\frac{1}{2}$,  we have

\be\label{eq:ffW2}
i{\mathcal M}(f^{\frac{1}{2}}\rightarrow f'^{\frac{1}{2}} W^+_L) 
&=& \frac{ig}{\sqrt{2}}\frac{1}{\sqrt{\zb}z} ( \frac{m_3^2}{m_W} z\zb-\frac{m_2^2}{m_W}z-2 m_W\zb) \n \\
 &=& i (- y_{f_1}m_1\sqrt{\zb}+y_{f_2}m_2\frac{1}{\sqrt{\zb}}- \frac{g}{\sqrt{2}} 2m_W\frac{\sqrt{\zb}}{z}) 
\ee

Based on the results above, it's also straightforwardly to work out the splitting amplitudes if the gauge boson couples to right-handed fermion current, i.e.   ${\mathcal L}=\frac{g}{\sqrt{2}}\bar{\psi}_1\gamma^{\mu}P_R\psi_2W_{\mu}$.   For $s_2=s_3=-\frac{1}{2}$, and $s_2=s_3=\frac{1}{2}$ respectively, the splitting amplitudes are 

\be\label{eq:ffW3}
i{\mathcal M}(f^{-\frac{1}{2}}\rightarrow f'^{-\frac{1}{2}} W^+_L) 
&=& \frac{ig}{\sqrt{2}}\frac{1}{\sqrt{\zb}z} (\frac{m_3^2}{m_W} z\zb-\frac{m_2^2}{m_W}z-2 m_W\zb)\n \\
i{\mathcal M}(f^{\frac{1}{2}}\rightarrow f'^{\frac{1}{2}} W^+_L) 
&=& \frac{ig}{\sqrt{2}}\frac{1}{\sqrt{\zb}z} (\frac{m_2^2}{m_W}z- \frac{m_3^2}{m_W} z\zb-2 m_W\zb) 
\ee

With splitting amplitudes for gauge boson coupling to left-handed current and right-handed current, we are able to calculate the splitting amplitudes  given by the Lagrangian  ${\mathcal L}=\frac{g}{\sqrt{2}}\bar{\psi}_1\gamma^{\mu}(Q_LP_L+Q_RP_R)\psi_2W_{\mu}$, with arbitrary $Q_L$ and $Q_R$.

\

\noindent {\large \bm{$W_L^+\rightarrow W^+_LW^0_L$} }

\

Following the Feynman rules in appendix \ref{sec:feynrules} and the momenta for $W_L^+\rightarrow W^+_LW^0_L$ are $k_3\rightarrow k_1 \  k_2$.  The splitting amplitude for  is given by the cubic vertex for vector bosons,

\be
i{\mathcal M}(W^+_L\rightarrow W^0_LW_L^+)=-ig& \{& [\epsilon_{n_1}(k_1)\cdot \epsilon_{n_2}(k_2)-\frac{i^2}{2}](-k_1+k_2)\cdot \epsilon_{n_3}(k_3)  \nonumber \\
                                      &+& [\epsilon_{n_1}(k_2)\cdot \epsilon_{n_3}(k_3)-\frac{i(-i)}{2}](-k_2-k_3)\cdot \epsilon_{n_1}(k_1) \nonumber \\
                                      &+& [\epsilon_{n_3}(k_3)\cdot \epsilon_{n_1}(k_1)-\frac{i(-i)}{2}](k_3+k_1)\cdot \epsilon_{n_2}(k_2) \} \nonumber 
\ee


$\epsilon_{n_i}\cdot \epsilon_{n_j}\sim \frac{m_W^2}{E^2}\theta$ is suppressed by both the factor of  $\frac{m_W}{2E_k}$ and $ \theta$,  so they are negligible.  Indeed, the simplest way is to choose $n_1=n_2=n_3=n$, which corresponds to the conventional light-cone gauge. This choice leads to $\epsilon_{n_i}\cdot \epsilon_{n_j}=0$.

The splitting amplitude then becomes

\be
i{\mathcal M}(W^+_L\rightarrow W^0_LW_L^+)= \frac{ig}{2}m_W \left[-\frac{(k_1-k_2)\cdot n}{n\cdot k_3} +\frac{(k_2+k_3)\cdot n}{n\cdot k_1}+\frac{-(k_3+k_1)\cdot n }{n\cdot k_2}\right]
\ee

 We also write $m_W=\frac{gv}{2}$,  plug all in.  After organization,  and making use of the definition of energy fraction $z/\zb$ in Eq.~(\ref{eq:z/zb}), the amplitude finally becomes 

\be
i{\mathcal M}_{W^+_L\rightarrow W^+_LW_L^0}=\frac{ig^2v}{2}\frac{z-\zb}{z\zb}(1+\frac{z\zb}{2})
\ee



\



\section{Conclusions}


In this paper we derived the Feynman rules of massive gauge theory in  physical gauges.   The model is $\theta_W\rightarrow 0$ limit of the standard model with gauge group $SU(2)_L$.  The main novelty is that we treat  gauge fields and goldstone fields uniformly as 5-component vector fields: $W^M=(W^{\mu}, \phi)$.  Making use of the new notation, we derived the propagator for vector bosons.   We noticed there is a remarkable similarity between massless gauge theory  and massive gauge  theory in the algebra level, making the derivation almost trivial.   We also derived the Feynman rules for vertices. Especially, we found that gauge-gauge-gauge vertex and goldstone-goldstone-gauge vertex can be combined into single W-W-W vertex with a common factor $\epsilon^{abc}$, which is obviously due to the remaining custodial symmetry in the scalar potential. 

We also investigated the structure of 3-point on-shell amplitudes.  We demonstrated that  all 3-point on-shell amplitudes -- W-W-W, h-W-W, $f$-$f'$-W -- satisfy on-shell gauge symmetry,  which is a reflection of on-shell gauge symmetry for general S-matrix.   This on-shell gauge symmetry ensures that amplitudes calculated with the new Feynman rules and with the usual Feynman rules are equivalent. We call this equivalence on-shell match condition.   Finally, making use of the new Feynman rules and on-shell match condition for 3-point amplitudes, we calculated some collinear splitting amplitudes in massive gauge theory.  

\

\

\

\noindent \textbf{Acknowledgements} \  \  \   The author thanks the discussions with Tao Han,  Brock Tweedie and Kaoro Hagiwara.  


\appendix

\section{Feynman Rules}
\label{sec:feynrules}

Convention:

\be
g^{\prime MN}=g^\prime_{MN}=\text{diag}(g_{\mu\nu}, -1/2) \  \  \  \  g^{MN}=g_{MN}=\text{diag}(g_{\mu\nu}, -1)\n
\ee 

\be
k^M = \left(\begin{array}{c}
k^{\mu} \\
-im_W 
\end{array}\right )   \  \  \  \   k^{*M} = \left(\begin{array}{c}
k^{\mu} \\
im_W 
\end{array}\right)     \  \  \  \   n^{M} = \left(\begin{array}{c}
n^{\mu} \\
0 
\end{array}\right)  \n   
\ee  

\be
n^2=0    \ \  \  \   \ \  k\cdot k^* = g_{MN}k^Mk^{*M} \n
\ee

\subsection*{Propagators}
\beqa
%
\vcenter{\begin{picture}(100,80)(0,0)
\SetColor{Black}\SetWidth{1.5}
\Text(53, 57)[]{$W^{\pm}$}
\Text(50,47)[]{$\longrightarrow$}
\Photon(10,40)(90, 40){3}{5}
\Text(0, 40)[]{$N$}
\Text(100, 40)[]{$M$}
\Text(50, 30)[]{$k$}
\end{picture}}
\hspace{-4.0in}
&=\ \ & \frac{-i}{k\cdot k^* +i\epsilon}
(g^{MN}-\frac{n^Mk^{* N}+k^Mn^{*N}}{n\cdot k-i\epsilon}+\xi \frac{k\cdot k^*}{(n\cdot k)^4}k^Mk^{* N})\nonumber  \\
\vcenter{\begin{picture}(100,80)(0,0)
\SetColor{Black}\SetWidth{1.5}
\Text(53, 57)[]{$W^0$}
\Text(50,47)[]{$\longrightarrow$}
\Photon(10,40)(90, 40){3}{5}
\Text(0, 40)[]{$N$}
\Text(100, 40)[]{$M$}
\Text(50, 30)[]{$k$}
\end{picture}}
\hspace{-4.0in}
&=\ \ & \frac{-i}{k\cdot k^* +i\epsilon}
(g^{MN}-\frac{n^Mk^{* N}+k^Mn^{*N}}{n\cdot k-i\epsilon }+\xi \frac{k\cdot k^*}{(n\cdot k)^4}k^Mk^{* N})\nonumber  \\
\vcenter{\begin{picture}(100,80)(0,0)
\SetColor{Black}\SetWidth{1.5}
\ArrowLine(10,40)(90, 40)
\Text(50, 30)[]{$k$}
\end{picture}}
\hspace{-4.0in}
&=\ \ & \frac{i(\slashed{k}+m_f)}{k^2-m_f^2+i\epsilon} \nonumber  \\
\vcenter{\begin{picture}(100,80)(0,0)
\SetColor{Black}\SetWidth{1.5}
\Text(53, 57)[]{$h$}
\DashLine(10,40)(90, 40){5}
\Text(50, 30)[]{$k$}
\end{picture}}
\hspace{-4.0in}
&=\ \ & \frac{i}{k^2-m_h^2+i\epsilon} \nonumber\\
\eeqa

\subsection*{Gauge-goldstone Sector}
\beqa
%
\vcenter{\begin{picture}(100,100)(0,0)
\SetColor{Black} \SetWidth{2}
\Photon(  10,25)(50,45){3}{3}
\Photon(50,45)(50,80){3}{3}
\Photon(50,45)(85, 25){3}{3} 
\Text(0,25)[]{$W_0^M$}
\Text(30,25)[]{\rotatebox{30}{$\longrightarrow$}}
\Text(30, 15)[]{$k_1$}
\Text(57,80)[l]{$W_-^N$}
\Text(40, 65)[]{\rotatebox{90}{$\longleftarrow$}}
\Text(31, 67)[]{$k_2$}
\Text(90, 25)[l]{$W_+^K$}
\Text(75, 40)[]{\rotatebox{-30}{$\longleftarrow$}}
\Text(80, 47)[]{$k_3$}
\end{picture}}
\hspace{-4.0in}
& =\ \ &    -ig \big(g^{\prime MN}(k_1-k_2)^{\rho}+g^{\prime NK}(k_2-k_3)^{\mu}+g^{\prime KM}(k_3-k_1)^{\nu}\big) \nonumber \\
\vcenter{\begin{picture}(100,100)(0,0)
\SetColor{Black} \SetWidth{2}
\Photon(17,17)(50,50){3}{3}
\Photon(17,83)(50,50){3}{3} 
\Photon(83,17)(50,50){3}{3}
\Photon(83, 83)(50,50){3}{3}
\Text(7,17)[]{$W_+^P$}
\Text(25,37)[]{\rotatebox{45}{$\longrightarrow$}}
\Text(95,17)[]{$W_-^\Sigma$}
\Text(75, 37)[]{\rotatebox{-45}{$\longleftarrow$}}
\Text(7,83)[]{$W_+^M$}
\Text(25,63)[]{\rotatebox{-45}{$\longrightarrow$}}
\Text(95,83)[]{$W_-^N$}
\Text(75, 63)[]{\rotatebox{45}{$\longleftarrow$}}
\end{picture}}
\hspace{-4.0in} 
&=&    ig^2 (2g^{\mu\rho}g^{\nu\sigma}-g^{\prime MN}g^{\prime P\Sigma}-g^{\prime M\Sigma}g^{\prime NP} ) -i(\lambda_h-\frac{g^2}{2})g^{M4}g^{N4}g^{P4}g^{\Sigma 4} \nonumber \\ 
\vcenter{\begin{picture}(100,100)(0,0)
\SetColor{Black} \SetWidth{2}
\Photon(17,17)(50,50){3}{3}
\Photon(17,83)(50,50){3}{3} 
\Photon(83,17)(50,50){3}{3}
\Photon(83, 83)(50,50){3}{3}
\Text(7,17)[]{$W_0^P$}
\Text(25,37)[]{\rotatebox{45}{$\longrightarrow$}}
\Text(95,17)[]{$W_0^\Sigma$}
\Text(75, 37)[]{\rotatebox{-45}{$\longleftarrow$}}
\Text(7,83)[]{$W_+^M$}
\Text(25,63)[]{\rotatebox{-45}{$\longrightarrow$}}
\Text(95,83)[]{$W_-^N$}
\Text(75, 63)[]{\rotatebox{45}{$\longleftarrow$}}
\end{picture}}
\hspace{-4.0in}
&=&    -ig^2 (2g^{\prime MN}g^{\prime P\Sigma}-g^{\prime MP}g^{\prime N\Sigma}-g^{\prime M\Sigma} g^{\prime PN})-i\frac{\lambda_h}{2} g^{M4}g^{N4}g^{P4}g^{\Sigma 4}    \nonumber \\
\vcenter{\begin{picture}(100,100)(0,0)
\SetColor{Black} \SetWidth{2}
\Photon(17,37)(50,60){3}{3}
\Photon(17,93)(50,60){3}{3} 
\Photon(83,27)(50,60){3}{3}
\Photon(83, 93)(50,60){3}{3}
\Text(7,27)[]{$W_0^P$}
\Text(25,47)[]{\rotatebox{45}{$\longrightarrow$}}
\Text(95,27)[]{$W_0^\Sigma$}
\Text(75, 47)[]{\rotatebox{-45}{$\longleftarrow$}}
\Text(7,93)[]{$W_0^M$}
\Text(25,73)[]{\rotatebox{-45}{$\longrightarrow$}}
\Text(95,93)[]{$W_0^N$}
\Text(75, 73)[]{\rotatebox{45}{$\longleftarrow$}}
\end{picture}}
\hspace{-4.0in}
&=&   
\includegraphics[width=12cm]{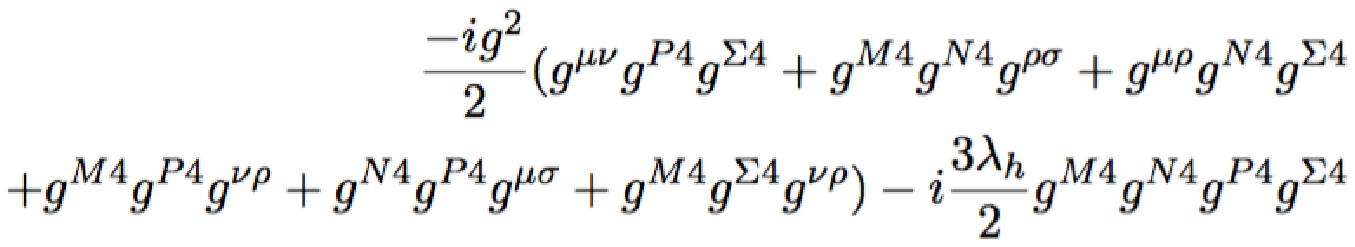} \n 
\eeqa

\vskip 0.5cm 

\subsection*{Higgs Sector and ``{\text VEV}" Sector}

\beqa
%
\vcenter{\begin{picture}(100,100)(0,0)
\SetColor{Black} \SetWidth{2}
\DashLine(  10,25)(50,45){3}
\Photon(50,45)(50,80){3}{3}
\Photon(50,45)(85, 25){3}{3} 
\Text(0,25)[]{$h$}
\Text(30,25)[]{\rotatebox{30}{$\longrightarrow$}}
\Text(30, 15)[]{$q$}
\Text(57,80)[l]{$W_-^M$}
\Text(40, 65)[]{\rotatebox{90}{$\longleftarrow$}}
\Text(31, 67)[]{$k_1$}
\Text(90, 25)[l]{$W_+^N$}
\Text(75, 40)[]{\rotatebox{-30}{$\longleftarrow$}}
\Text(80, 47)[]{$k_2$}
\end{picture}}
\hspace{-4.0in}
& =\ \ &    -\frac{g}{2} \left(g^{ N4}(k_1^{\mu}-q^{\mu})+g^{ M4}(k_2^{\nu}-q^{\nu})\right)+igm_Wg^{\mu\nu}-i\frac{\lambda_hv}{2}g^{M4}g^{N4} \nonumber \\
\vcenter{\begin{picture}(100,100)(0,0)
\SetColor{Black} \SetWidth{2}
\DashLine( 10,25)(50,45){3}
\Photon(50,45)(50,80){3}{3}
\Photon(50,45)(85, 25){3}{3} 
\Text(0,25)[]{$h$}
\Text(30,25)[]{\rotatebox{30}{$\longrightarrow$}}
\Text(30, 15)[]{$q$}
\Text(57,80)[l]{$W_0^M$}
\Text(40, 65)[]{\rotatebox{90}{$\longleftarrow$}}
\Text(31, 67)[]{$k_1$}
\Text(90, 25)[l]{$W_0^N$}
\Text(75, 40)[]{\rotatebox{-30}{$\longleftarrow$}}
\Text(80, 47)[]{$k_2$}
\end{picture}}
\hspace{-4.0in}
& =\ \ &    -\frac{g}{2} \big((k_1-q)^{\mu}g^{ N4}+g^{ M4}(k_2-q)^{\nu}\big)+igm_Wg^{\mu\nu}-i\frac{\lambda_hv}{2}g^{M4}g^{N4} \nonumber \\
\vcenter{\begin{picture}(100,100)(0,0)
\SetColor{Black} \SetWidth{2}
\DashLine(17,17)(50,50){5}
\DashLine(17,83)(50,50){5} 
\Photon(83,17)(50,50){3}{3}
\Photon(83, 83)(50,50){3}{3}
\Text(7,17)[]{$h$}
\Text(25,37)[]{\rotatebox{45}{$\longrightarrow$}}
\Text(95,17)[]{$W_+^\Sigma$}
\Text(75, 37)[]{\rotatebox{-45}{$\longleftarrow$}}
\Text(7,83)[]{$h$}
\Text(25,63)[]{\rotatebox{-45}{$\longrightarrow$}}
\Text(95,83)[]{$W_-^N$}
\Text(75, 63)[]{\rotatebox{45}{$\longleftarrow$}}
\end{picture}}
\hspace{-4.0in}
& =\ \ &   -i\frac{g^2}{2} g^{ \nu\sigma} -i\frac{\lambda_h}{2}g^{N4}g^{\Sigma 4}   \nonumber \\
\vcenter{\begin{picture}(100,100)(0,0)
\SetColor{Black} \SetWidth{2}
\DashLine(17,17)(50,50){5}
\DashLine(17,83)(50,50){5} 
\Photon(83,17)(50,50){3}{3}
\Photon(83, 83)(50,50){3}{3}
\Text(7,17)[]{$h$}
\Text(25,37)[]{\rotatebox{45}{$\longrightarrow$}}
\Text(95,17)[]{$W_0^\Sigma$}
\Text(75, 37)[]{\rotatebox{-45}{$\longleftarrow$}}
\Text(7,83)[]{$h$}
\Text(25,63)[]{\rotatebox{-45}{$\longrightarrow$}}
\Text(95,83)[]{$W_0^N$}
\Text(75, 63)[]{\rotatebox{45}{$\longleftarrow$}}
\end{picture}}
\hspace{-4.0in}
& =\ \ &   -i\frac{g^2}{2} g^{\nu\sigma} -i\frac{\lambda_h}{2}g^{N4}g^{\Sigma 4}   \nonumber  \\
\vcenter{\begin{picture}(100,80)(0,0)
\SetColor{Black} \SetWidth{2}
\DashLine(  10,5)(50,25){6}
\Text(50,25)[]{$$}
\DashLine(50,25)(50,60){6}
\DashLine(85, 5)(50,25){6}
\Text(0,5)[]{$h$}
\Text(57,60)[l]{$h$}
\Text(90, 5)[l]{$h$}
\end{picture}}
\hspace{-4.0in}
& =\ \ &   -i \frac{3\lambda_h v}{2}  \nonumber  \\
\vcenter{\begin{picture}(100,95)(0,0)
\SetColor{Black} \SetWidth{2}
\DashLine(17,7)(50,40){5}
\DashLine(17,73)(50,40){5} 
\DashLine(83,7)(50,40){5}
\DashLine(83, 73)(50,40){5}
\Text(7,7)[]{$h$}
\Text(95,7)[]{$h$}
\Text(7,73)[]{$h$}
\Text(95,73)[]{$h$}
\end{picture}}
\hspace{-4.0in}
& =\ \ &    -i\frac{3}{2}\lambda_h \nonumber  
\eeqa

\vskip 0.5cm

\subsection*{Fermion Sector}

\beqa
\vcenter{\begin{picture}(100,95)(0,0)
\SetColor{Black} \SetWidth{2}
\Text(0,5)[]{$d_i$}
\ArrowLine(  10,5)(50,25)
\Photon(50,25)(50,60){3}{3}
\ArrowLine( 50,25)(85,5)
\Text(90,5)[l]{$u_j$}
\Text(57,60)[l]{$W_{+}^M$}
\end{picture}}
\hspace{-4.0in}
& =\ \ & (-i \frac{g}{\sqrt{2}} \gamma^{\mu}P_L \nonumber - (y_dP_R-y_uP_L)g^{M4} ) V_{ij} \\
\vcenter{\begin{picture}(100,95)(0,0)
\SetColor{Black} \SetWidth{2}
\Text(0,5)[]{$u_i$}
\ArrowLine(  10,5)(50,25)
\Photon(50,25)(50,60){3}{3}
\ArrowLine( 50,25)(85,5)
\Text(90,5)[l]{$d_j$}
\Text(57,60)[l]{$W_{-}^M$}
\end{picture}}
\hspace{-4.0in}
& =\ \ & (-i \frac{g}{\sqrt{2}} \gamma^{\mu}P_L \nonumber - (y_uP_R-y_dP_L)g^{M4} ) V^*_{ij}\\
%
\vcenter{\begin{picture}(100,80)(0,0)
\SetColor{Black} \SetWidth{2}
\Text(0,5)[]{$f$}
\ArrowLine(  10,5)(50,25)
\Photon(50,25)(50,60){3}{3}
\ArrowLine( 50,25)(85,5)
\Text(90,5)[l]{$f$}
\Text(57,60)[l]{$W_0^N$}
\end{picture}}
\hspace{-4.0in}
& =\ \ &  -ig \gamma^{\mu} \left(T^3_f P_L \right)-y_f\gamma_5T_3g^{M4} \nonumber 
\eeqa



 
 



 
 

\end{document}